# Temporal Evolution of Bradford Curves in Specialized Library Contexts

Haobai Xue[1], Xian Liu[2]

(1. xuehb@utszlib.edu.cn; Shenzhen Science & Technology Library/University Town Library of Shenzhen, 2239 Lishui Road, Nanshan District, Shenzhen 518055, China;

2. xliu@shenzhong.net; Shenzhen Middle School, 1068 Nigang West Road, Luohu District, Shenzhen, Shenzhen 518055, China)

**Abstract:**

Bradford's law of bibliographic scattering is a fundamental principle in bibliometrics, offering valuable guidance for academic libraries in literature search and procurement. However, Bradford curves can exhibit various shapes over time, and predicting these shapes remains a challenge due to a lack of causal explanation. This paper attributes the deviations from the theoretical J-shape to integer constraints on the number of journals and articles, extending Leimkuhler and Egghe's formulas to encompass highly productive core journals, where the theoretical journal number falls below one. Using the Simon-Yule model, key parameters of the extended formulas are identified and analyzed. The paper explains the reasons for the Groos Droop and examines the critical points for shape changes. The proposed formulas are validated with empirical data from literature, demonstrating that this method can effectively predict the evolution of Bradford curves, thereby aiding academic libraries in the procurement and utilization of scientific literature.

## 1 Introduction

### 1.1 Introduction

As one of the three fundamental laws of bibliometrics, Bradford's law of bibliographic scattering has numerous potential applications in academic libraries. For instance, it can help identify the core journals or publishers in a specific research area, guiding librarians in journal and book procurement decisions (Barrantes et al., 2023). Additionally, it can assist readers and librarians in literature searches by quickly pinpointing key WoS research areas or IPC classes related to a topic (Sheikh et al., 2022). However, preparing Bradford curves can be time-consuming, particularly for journals with only one or two relevant papers. Moreover, since the scientific literature in a certain discipline or research area often increases exponentially or passes through various developmental stages (Larivière et al., 2008), a Bradford curve prepared at any given time may not be directly applicable many years later without adjustment. Some mathematical formulas can predict the shape of Bradford curve, but they typically result in a J-shaped curve. In practice, Bradford's curve can take at least six different shapes, with the notable S-shaped curve with the so-called Groos Droop (Groos, 1967). There is a lack of causal explanations for this bibliometric law and comprehensive empirical examples (Wagner-Döbler, 1997). Therefore, predicting the evolution of Bradford curve remains an open question that warrants further investigation.

This paper attributes the different shapes of Bradford curve to the integer constraints of journal and paper numbers. If journal productivity $n$ is high enough that the corresponding theoretical journal number $f_t(n) = C/n^\alpha$ falls below one, the actual journal number $f_e(n)$ can only be zero or one. Consequently, the discrete nature of journal numbers causes the core zone to deviate from the theoretical results of the Lotka or Simon-Yule model. To address this issue, this paper proposes two different formulas for the core zone and the normal zone. Key parameters of these formulas are identified and studied through theoretical analysis and Monte Carlo simulation of the Simon-Yule model. The causes of the Groos Droop are explained, and the critical points for shape changes are examined. Finally, the proposed formulas are validated using empirical data from the literature. The findings suggest that the proposed method can predict the evolution of Bradford's curves, thereby guiding academic libraries in the procurement and utilization of scientific literature.

**1.2 Literature Review**

Bradford's law was first proposed by Bradford in 1934 (Bradford, 1934) but did not gain wide recognition until Vickery further developed the theory in 1948 (Vickery, 1948). According to Bradford's law, if journals are arranged in descending order of productivity and divided into $p$ groups with the same number of papers, the number of journals in each group $n_i$ follows the ratio $n_1: n_2: \cdots : n_p = 1: k: \cdots : k^{p-1}$, where $k$ is the Bradford multiplier. Besides this verbal form, Bradford's law can also be depicted as a J-shaped curve by plotting the accumulated productivity $R(r)$ of the first $r$ journals against the natural logarithm of the journal rank $r$. Leimkuhler proposed the mathematical formula for Bradford's curve in 1967 (Leimkuhler, 1967), and Egghe developed a method for determining the parameters of this formula in 1990 (Egghe, 1990). In Egghe's formula, $R(r) = a\log(1 + br)$, where the key parameters $a$ and $b$ can be calculated from the article number $A$, journal number $T$, and the productivity $y_m$ of the most productive journal. Although Egghe's formula matches well with many bibliographies, it corresponds to a J-shaped curve, which deviates from those with a Groos Droop (Egghe, 1990).

Incomplete bibliographies were initially believed to cause the Groos Droop, but further research refuted this hypothesis (Qiu & Tague, 1990). Egghe demonstrated that if the ranking of each journal $r$ is transformed into $r' = r + r_0$ by adding a large constant $r_0 > 1/b$, then new curve will concave downwards, showing a Groos Droop (Egghe & Rousseau, 1988). The merging of different bibliographies, each with a different maximum journal productivity $y_m^{(i)}$, could explain the large constant $r_0$ (Egghe & Rousseau, 1988). However, it is also likely that the large core regions (regions with the most productive journals where $f_t(n_i) < 1$) of some bibliographies contribute to the large $r_0$ (Chen & Leimkuhler, 1987). Essentially, $y_m$ in Egghe's formula denotes the journal productivity where $f_t(y_m) = C/y_m^\alpha \approx 1$ (Egghe, 1985), rather than the maximum yield $X_1$ of a journal as claimed by Egghe himself. Thus, if the total number of these journals $T_0$ exceeds the critical value $r_0 = 1/b$, a Groos Droop will emerge. This paper adopts this explanation and extends Egghe's formula to predict the evolution of Bradford's curves.

In the 1990s, research interest in Bradford's law shifted from the static presentation of data at a particular time to its dynamic and evolutionary aspects (Oluić-Vuković, 1998). Oluić-Vuković studied how the increase in productivity of core journals affected the shape of the distribution curve over time (Oluić-Vuković, 1989). By analyzing the research output of Croatian scholars in different subjects, she concluded that the Groos Droop or S-shaped curve is caused by an increase in the concentration/dispersal disparity, reflected by the rise in the core/periphery ratio (Oluić-Vuković, 1991). The dynamic evolution of Bradford curves and the emergence of the Groos Droop are presented in her 1992 study (Oluić-Vuković, 1992), and other similar empirical studies partitioning bibliographies over time were conducted by Garg (Garg et al., 1993), Wagner-Döbler (Wagner-Döbler, 1997) and Sen (Sen & Chatterjee, 1998).

Meanwhile, stochastic models like the Simon-Yule model have increasingly been used to study the dynamic characteristics of bibliometric laws (Oluić‐Vuković, 1997, 1998). Initially introduced by Yule in 1924 for studying the distribution of biological genera by species number, the Simon-Yule model gained recognition when Simon expanded it in 1955 to analyze the frequency distributions of words in writing samples (Simon, 1955). Besides employing theoretical methods for precisely solving the constant entry rate $\alpha$ of new sources (Simon, 1955), Monte Carlo simulations have been used to explore more complex scenarios, such as declining entry rates $\alpha_t$ (Simon & Van Wormer, 1963) and autocorrelated growth rates $\gamma$ of established journals (also referred to as aging or obsolescence rate) (Ijiri & Simon, 1977). Chen et al. (Chen, 1989; Chen et al., 1994; Chen et al., 1995) first used the Simon-Yule model to study the evolution of Lotka's and Bradford's laws over time. They found the entry rate $\alpha_t$ and the autocorrelated growth rate $\gamma$ have significant yet opposite effect on the Bradford curves, offering an explanation for the various types of Bradford curves (Chen et al., 1995).

Later, Oluić-Vuković also explored the dynamics of Bradford distribution using the Simon-Yule model but found its steady-state solution too restricted to handle time variations, limiting its applicability (Oluić‐Vuković, 1997, 1998). This paper also utilizes the Simon-Yule model to examine different scenarios' effects on key parameters (e.g., journal number $T_0$, article number $A_0$ and maximum productivity $X_1$ of the core region) of the extended Egghe's formula. However, it is not used directly to forecast the evolution of Bradford curves or to compare them with empirical data. Instead, key parameters are estimated from past empirical data to improve predictions of Bradford curve evolution in the future.

## 2 Theoretical Study

### 2.1 Simon-Yule Model

The Simon's generating mechanism for the Bradford distribution is based on the following two assumptions, where the $f_t(n, t)$ denotes the number of journals that have published exactly $n$ papers in the first $t$ published papers.

**Assumption I:** There is a constant probability $\alpha$ that the $(t + 1)$-th paper is published in a new journal – a journal that has not published in the first $t$ papers;

**Assumption II:** The probability that the $(t + 1)$-th paper is published in a journal that has published $n$ papers is proportional to $nf(n, t)$ – that is, to the total numbers of papers of all journals that have published exactly $n$ papers.

Therefore, if there are $A$ papers at a given time, then the corresponding journal number $T$ is approximately $T = A\alpha$. Based on Simon's two assumptions, the steady-state solution of the Bradford distribution can be written as (Chen, 1989):

$$f_t(n) = \rho B(n, \rho + 1) \approx \rho \Gamma(\rho + 1) n^{-(\rho+1)} \qquad (1)$$

where $B$ is the beta function, $\Gamma$ is the gamma function, and $\rho$ is a function of the entry rate of new journal $\alpha$, defined as $\rho = 1/(1 - \alpha)$. Equation (1) suggests that the analytical outcome of the Simon-Yule model aligns with the Lotka's law when $\rho \approx 1$.

In addition to the analytical solutions, Monte Carlo simulations are conducted for $\alpha = 0.15$, and the results are compared with the theoretical results of Equation (1), as shown in Figure 1. Detailed procedures for these simulations can be found in (Simon & Van Wormer, 1963) and are thus omitted here. To reduce inherent randomness, each case is simulated $N = 10^4$ times, utilizing only the medians of these simulations as the final outputs.

Figure 1 illustrates distinct zones in simulation results: a normal zone (blue circles) and a core zone (red squares). This distinction arises from the necessity for actual journal numbers $f_e(n)$ to be integers, unable to fall below one. Thus, when the journal productivity $n$ is high enough for theoretical journal number $f_t(n)$ to drop below one, actual journal numbers $f_e(n)$ are constrained to zone or one, deviating from theoretical predictions as shown by the red squares and black dots in Figure 1. Moreover, despite the smaller number of journals in the core region, their contribution to paper count is significant, as shown in Figure 1(b). Hence, accurate prediction of paper count for each journal $X_r$ in the core zone is crucial for depicting the Bradford curve faithfully.

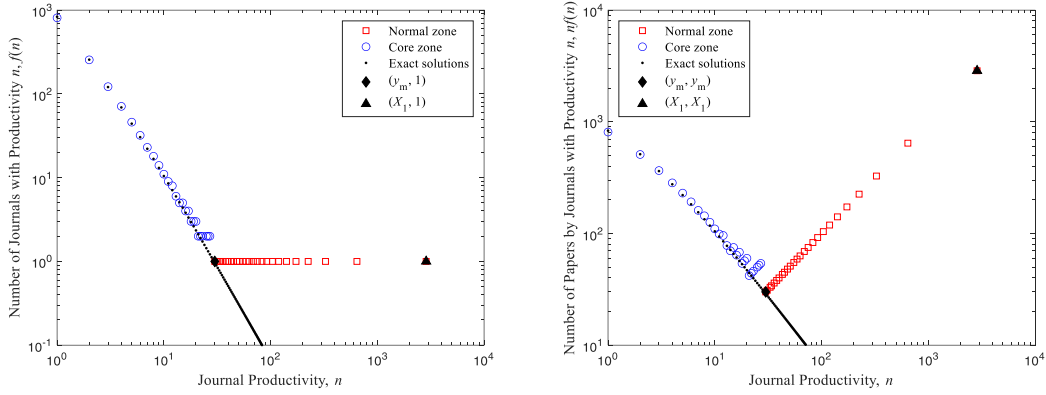

Figure 1 Comparisons of the theoretical and numerical results: (a) number of journals $f(n)$ with productivity $n$; (b) number of papers $nf(n)$ produced by journals with productivity $n$.

Estimating journal productivity $X_r$ of the core region involves determining journal count $T_0$ and paper count $A_0$ first. Since Figure 1 shows a close match between journal number $f(n)$ and paper number $nf(n)$ in the normal region, total journal count $T_1$ and total paper count $A_1$ of the normal region can be directly obtained by summing all journal and paper numbers. These are calculated as $T_1 = \sum_{n=1}^{y_m} f(n)$ and $A_1 = \sum_{n=1}^{y_m} nf(n)$, where $y_m$ is the journal productivity when $f_t(y_m) \approx 1$. In this context, $y_m$ can be seen as the productivity of the most productive journal in the normal region, as indicated by the black diamond in Figure 1. According to Equation (1), the analytical expression for $y_m$ can be derived as:

$$y_m = [A(\rho - 1)\Gamma(\rho + 1)]^{\frac{1}{\rho+1}} \qquad (2)$$

Once $y_m$ is calculated, total journal count $T_0$ and paper count $A_0$ of the core region can be calculated as $T_0 = T - T_1$ and $A_0 = A - A_1$. Alternatively, they can be directly calculated as:

$$T_0 \approx \int_{y_m}^{+\infty} Tf(n)\,\mathrm{d}n = \frac{y_m}{\rho} \qquad (3)$$

$$A_0 \approx \int_{y_m}^{+\infty} Tnf(n)\,dn = \frac{y_m^2}{\rho - 1} \quad (4)$$

where $y_m$ is calculated using Equation (2).

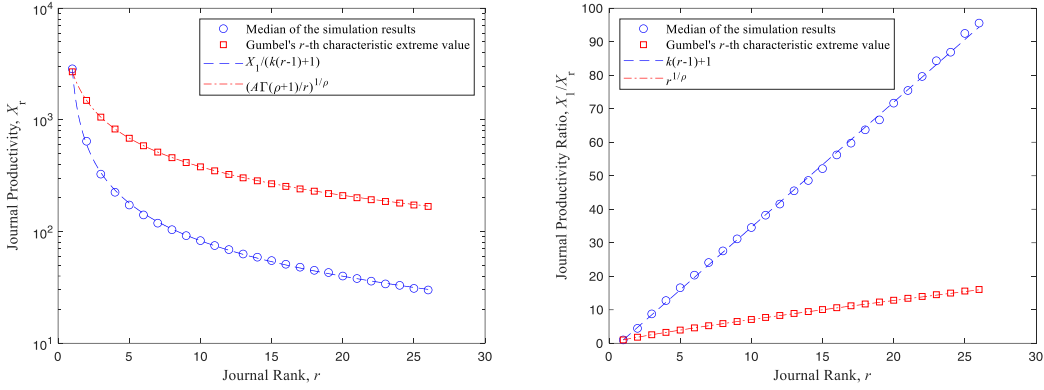

Figure 2 Journal productivity in the core region $X_r$ as a functions of journal rank $r$: (a) Journal productivity $X_r$ as a function of $r$; (b) Journal productivity ratio $X_1/X_r$ as a function of $r$.

In the Simon-Yule model with constant entry rate $\alpha$, the maximum number of papers one journal can have, $X_1$, can be estimated using Gumbel's $r$-th characteristic extreme theory (Glänzel, 2010, 2013):

$$G(X_r) \approx \int_{X_r}^{+\infty} f(i)\,di = \frac{r}{A} \quad (5)$$

By solving this equation, it can be derived that the productivity of the most productive journal, $X_1$, can be expressed as:

$$X_1 = [A\Gamma(\rho + 1)]^{\frac{1}{\rho}} = (\rho - 1)^{-\frac{1}{\rho}} y_m^{\frac{\rho+1}{\rho}} \quad (6)$$

The productivity of the $r$-th most productivity journal, $X_r$, is related to $X_1$ by $X_r = X_1 r^{-1/\rho}$. Figure 2 compares Gumbel's $r$-th characteristic extreme values with the medians of the simulation results. While Gumbel's theory effectively predicts the largest paper number $X_1$, it falls short in estimating other paper numbers in the core zone $X_r$, for $r = 2, 3, \cdots, T_0$. Thus, an alternative method is assumed in this paper, where all other $X_r$, for $r = 2, 3, \cdots, T_0$, are related to the largest paper number $X_1$ through the equation:

$$\frac{X_1}{X_r} = k(r - 1) + 1 \quad (7)$$

where $k$ is the only parameter waiting to be determined. This equation can be derived from the Equation (8) of Reference (Chen, 1989), by assuming $(r_i - r_1)/(r_1 + b)$ and $-1/c$ are relatively small. The validity of this equation is also supported by Figure 2(b), where the blue circles represent the simulation results, and the blue dashed lines show the linear fitting results. Hence, the productivity of the $r$-th most productive journal can be derived from Equation (7), and the cumulative productivity of the first $r$ most productive journals can be written as:

$$R_c(r) = \sum_{i=1}^{r} \frac{X_1}{k(i-1)+1} \tag{8}$$

If there are $T_0$ journals with $A_0$ papers in the core region and the numbers of $T_0$ and $A_0$ are known, then the parameter $k$ can be calculated from the equation $R_c(T_0) = A_0$. Then, Equation (8) can be used to predict the evolution of the core regions ($r \leq T_0$) of Bradford's curves.

**2.2 Egghe's formula**

After removing the $T_0$ journals and $A_0$ papers of the core region, the remaining $T_1$ journals and $A_1$ papers align well with the theoretical results predicted by Equation (1). Consequently, they follow the Lotka's law, and their Bradford curve can be predicted using the revised Leimkuhler and Egghe's formula (Egghe, 1990):

$$R(r_1) = a\log(1 + br_1) \tag{9}$$

where the key parameters $a$ and $b$ are defined as:

$$a = \frac{A_1}{\log(e^\gamma y_m)} \tag{10}$$

$$b = \frac{e^\gamma y_m - 1}{T_1} \tag{11}$$

where $\gamma$ is Euler's constant, $\gamma \approx 0.5772$, and $y_m$ is the journal productivity when the corresponding theoretical journal number $f_t(y_m) \approx 1$. While $y_m$ can be directly calculated from Equation (2), it can also be estimated using the following equation if the values of $X_1$, $T_0$ and $A_0$ are known:

$$y_m \approx \frac{X_1}{k(T_0 - 1) + 1} \tag{12}$$

Since the core region's journal productivity is higher than the normal region, these significant journals rank lower. Consequently, the Bradford curve for the normal region starts at the point $(T_0, A_0)$. Each rank $r_1$ in the normal region should be transformed to $r = r_1 + T_0$, and the cumulative productivity of the first $r$ journals $R(r)$ should be transformed into $R_n(r) = R(r_1) + A_0$. The revised Egghe's formula for the normal region is then written as:

$$R_n(r) = R(r - T_0) + A_0 = a\log[1 + b(r - T_0)] + A_0 \tag{13}$$

Equation (13) can be used to predict the dynamic evolution of the normal regions ($T_0 < r < T$) of the Bradford's curve. Therefore, Equations (8) and (13) together can be used to predict the dynamic evolution of the Bradford's curves.

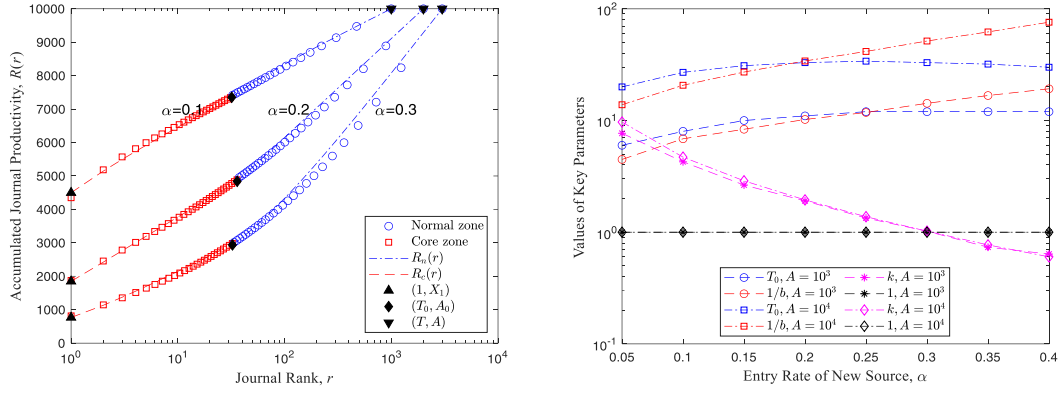

Figure 3 the evolution of the Bradford curve and the cause of the Groos Droop: (a) the evolution of the Bradford curve; (b) the cause of the Groos Droop

Figure 3(a) displays the Bradford curve. The blue circles represent the normal zone, while the red squares indicate the core zone. The blue dashed lines show the prediction results of Equation (13), and the red dotted lines show the prediction results of Equation (8). The black upper triangle, diamond and lower triangle represent the points $(1, X_1)$, $(T_0, A_0)$ and the $(T, A)$ respectively. From the discussion above, it is evident that these three points and two lines are crucial for predicting the evolution of the Bradford curve.

**2.3 Groos Droop**

Groos was the first to observe that in some datasets, when the journal productivity is low, the Bradford curve tend to bend downward (Groos, 1967). Egghe (Egghe & Rousseau, 1988) explained the cause of the Groos Droop as merging datasets. However, this section demonstrates that the core region's existence causes the Groos Droop in the normal region.

The first and second derivatives of $R_c(r)$ for the core region can be derived from Equation (8):

$$\frac{\partial R_c(r)}{\partial (\log r)} = \frac{X_1 r}{k(r-1)+1} \tag{14}$$

$$\frac{\partial^2 R_c(r)}{\partial (\log r)^2} = \frac{X_1(1-k)r}{[k(r-1)+1]^2} \tag{15}$$

From Equation (15), it can be noted that when $k > 1$, $\frac{\partial^2 R_c(r)}{\partial (\log r)^2} < 0$, causing the Bradford curve for the core region to concave downward. Conversely, when $k < 1$, $\frac{\partial^2 R_c(r)}{\partial (\log r)^2} > 0$, causing the curve to concave upward. As the entry rate of new journals $\alpha$ increases, the number of journals $T$ rises, and the distribution of articles become more dispersed, leading to a decrease in the largest journal productivity $X_1$. From Equation (8) and $R_c(T_0) = A_0$, we know that a lower $X_1$ results in a lower $k$ if $A_0$ is relatively constant. Therefore, as $\alpha$ increases, the Bradford curve for the core region will gradually concave upward, as shown in Figure 3.

Similarly, the first and second derivatives of $R_n(r)$ for the normal region can be derived from Equation (13):

$$\frac{\partial R_n(r)}{\partial (\log r)} = \frac{abr}{b(r-T_0)+1} \tag{16}$$

$$\frac{\partial^2 R_n(r)}{\partial (\log r)^2} = \frac{ab(1-bT_0)r}{[b(r-T_0)+1]^2} \tag{17}$$

From Equation (17), it can be noted that when $T_0 > 1/b$, $\frac{\partial^2 R_c(r)}{\partial (\log r)^2} < 0$, the Bradford's curve for the normal region will concave downward, showing a Groos Droop. When $T_0 < 1/b$, $\frac{\partial^2 R_c(r)}{\partial (\log r)^2} > 0$, the curve will concave upwards, forming a J-shaped curve. As the entry rate of new journals $\alpha$ increases, Figure 3(a) shows that the $T_0$ will eventually fall below $1/b$, causing the Bradford curve for the normal region to concave upward, similar to the core region.

Figure 3 (b) illustrates the variation of key parameters $T_0$, $1/b$ and $k$ with the entry rate $\alpha$. When $A = 10^4$, the normal region will start to concave upward at critical point $\alpha_n \approx 0.2$, while the core region will do so at $\alpha_c \approx 0.3$. Therefore, when $\alpha < 0.2$, the entire Bradford curve will concave downward; when $\alpha > 0.3$, it will concave upward. For $0.2 < \alpha < 0.3$, the Bradford curve will exhibit a reversed S shape, with the core region concaving downward and the normal region concaving upward. Figure 3(a) shows the three shapes of Bradford curves. In this specific case, since $\alpha_n < \alpha_c$, there is no S-shaped Bradford curve. This is because the aging of journals is not considered here, making the largest journal productivity $X_1$ relatively large. When the aging effects are considered, which will be discussed in detail in Section 3.2, $X_1$ decreases significantly, which results in a lower $k$ and thus a much lower $\alpha_c$. If $\alpha_c < \alpha_n$, an S-shaped Bradford curve will appear for $\alpha_c < \alpha < \alpha_n$, with the core region concaving upward and the normal region concaving downward.

**2.4 Bradford Dynamics**

Given the analytical expression of $T_0$, $A_0$ and $X_1$ (Equations (3), (4), and (6)), the core region of the Bradford curve at any time can be predicted using Equation (8). The parameters for the normal regions can be derived from these factors through $T_1 = T - T_0$, $A_1 = A - A_0$ and Equation (12), allowing the normal region of the Bradford curve to be predicted using Equation (13).

The Bradford's curves for the constant entry rate scenario are shown in Figure 4(a), Here, the red squares and blue circles represent the simulation results for the core and normal regions, respectively, while the red dashed lines and blue dotted lines represent the theoretical results of Equations (8) and (13). The key points $(1, X_1)$, $(T_0, A_0)$ and $(T, A)$ are also shown as black upper triangle, diamond and lower triangle in Figure 4(a). It is notable that the although the core region contains far fewer journals than the normal region, its representation is significant due to the x-axis's log scale. Consequently, journals with lower ranks are better represented in Figure 4(a).

Figure 3 shows that when $\alpha = 0.1$, the entire Bradford curve concaves downward for $10^3 < A < 10^4$. Figure 4(a) depicts the evolution of Bradford curves as the paper number $A$ increases from $10^3$ to $10^4$, aligning well with theoretical predictions.

Figure 4(b) presents the simulation and analytical results for $T_0$, $A_0$ and $X_1$. Hollow symbols represent the simulation results, while solid symbols denote the analytical ones. It can be observed that these three key factors are linear functions of the paper number $A$, as indicated by Equation (18):

$$\log(Y) = a_\rho + b_\rho \log(A) \tag{18}$$

where the constant $a_\rho$ and $b_\rho$ are functions of $\rho$. Since $\rho$ is approximately one, $a_\rho$ and $b_\rho$ can be considered constants. The close match between analytical and numerical results confirms the validity of Equations (3), (4), and (6). The analytical result for $T_0$ is slightly lower than the numerical ones because, in the numerical results, all journals with only one relevant paper are considered part of the core region, whereas, in theory, some of them belong to the normal region.

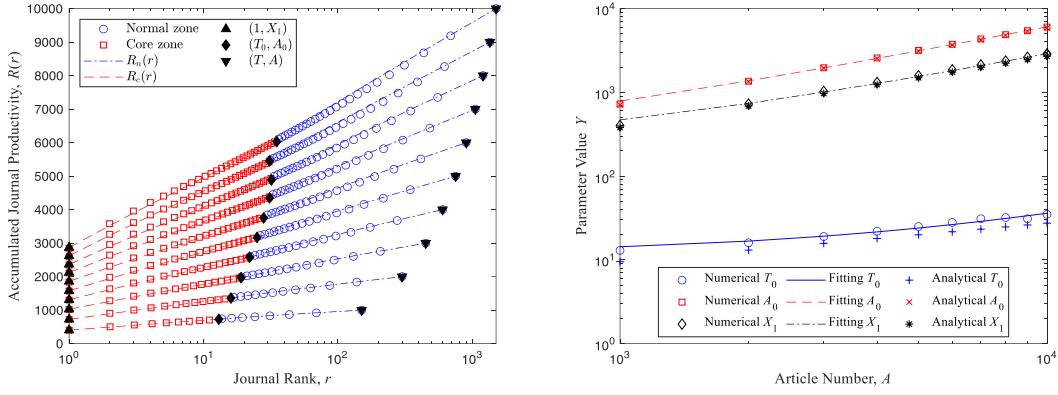

figure 4 the dynamics of the Bradford curve and the variation of key parameters when $\alpha = 0.15$. (a) the dynamics of the Bradford curves; (b) the variations of key parameters.

## 3 Numerical Study

### 3.1 Decreasing Entry Rate

Assume the probability of adding a new journal decreases linearly over the time.

$$\alpha(t) = \alpha_s - kt \tag{19}$$

where $k$ is a constant, $k = (\alpha_s - \alpha_f)/A_f$, where $\alpha_f$ and $A_f$ are the entry rate of new journals and the total number of articles in the final state, respectively. The accumulated number of journals can then be expressed as:

$$T = \sum_{t=1}^{A} \alpha(t) = \alpha_s A - \frac{1}{2}kA^2 \tag{20}$$

Using Equation (20), a quadratic fitting of $T$ and $A$ allows us to determine the values of $\alpha_s$ and $\alpha_f$. Once these are known, the average entry rate $\bar{\alpha} = (\alpha_s + \alpha_f)/2$ can be used to calculate the analytical results.

Figure 5 shows the dynamic evolution of Bradford curves and the variations of key parameters when the entry rate decreases linearly from 0.2 to 0.1. The proposed method effectively predicts these variations, with analytical results using $\bar{\alpha}$ matching well with the simulation results. Although the numerical results for $A_0$ and $X_1$ are slightly lower than the analytical ones, this suggests that a decreasing entry rate has a slight negative impact on their increase. However, the effect on the overall shape of the Bradford's curve and key parameters is relatively insignificant, indicating that the analytical results for a constant entry rate $\bar{\alpha}$ can still be used to predict key parameters without significant errors.

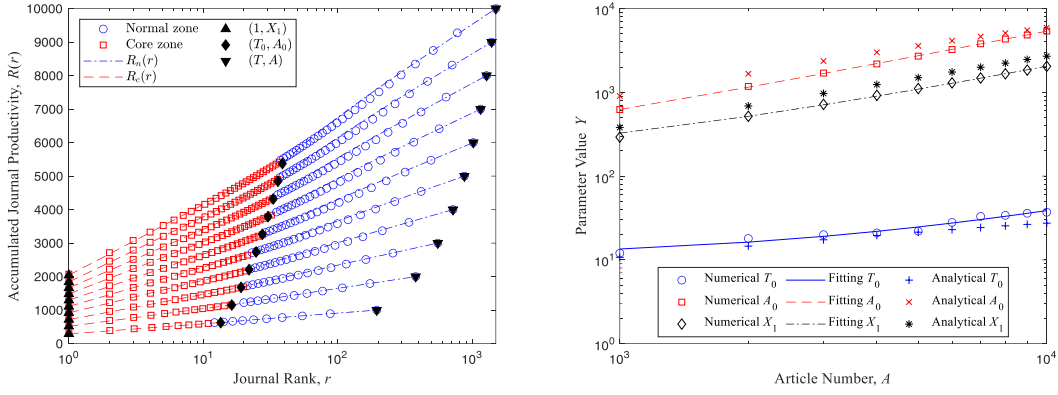

Figure 5 the dynamics of the Bradford curve and the variation of key parameters when $\alpha$ decreases linearly from 0.2 to 0.1. (a) the dynamics of the Bradford curves; (b) the variations of key parameters.

**3.2 Aging Rate of Journals**

Simon assumes that only one paper gets published in each time period, and models the probability of a journal increasing in paper number as proportional to a weighted sum of its past increments. These increments are weighted by a factor that decreases geometrically over time, with the rate of decrease denoted as $\gamma$.

Let $y_j(k)$ represent the change in paper number of the $j$-th journal during the $k$-th time interval, where $y_j(k)$ is either 1 (indicating a unit increment) or 0 (indicating no change). The paper number of the $j$-th journal at the end of the $k$-th interval is given by $\sum_{\tau=1}^{k} y_j(\tau)$ The expected increment in paper number during the $(k+1)$-th interval is:

$$p[y_j(k+1) = 1] = \frac{1}{W_k} \sum_{\tau=1}^{k} y_j(\tau) \gamma^{k-\tau} \tag{21}$$

where $W_k$ is a time-dependent function consistent across all journals, defined as $W_k = \sum_{j=1}^{T} w_j(k)$ with $w_j(k) = \sum_{\tau=1}^{k} y_j(\tau) \gamma^{k-\tau}$. The parameter $\gamma$ determines how quickly the influence of past growth diminished and is thus referred to as the aging rate of journals in this paper.

Figure 6 illustrates the impact of the aging rate of journals $\gamma$ on the dynamics of Bradford's curves and key parameters. It shows that the aging factor increases $T_0$, making the normal region more concave downward. The aging effect also markedly reduces $X_1$ by weakening the Mathew effect, where successful journals attract more papers. As older journals lose their appeal, this "success breeds success" effect diminished, leading to a substantial decrease in $X_1$, as shown in Figure 6(a). Consequently, while the number of articles $A_0$ in the core zone remains relatively unchanged, $T_0$ must increase to offset the reduction in $X_1$. This expansion of the core region increases its share of the Bradford curve, shaping it into a J-shape due to the reduction of $k$ and $X_1$, as discussed in Section 2.3. Additionally, with the increase in $T_0$, it is more likely that $T_0$ will exceed $1/b$, further contributing to the concave downward shape of the normal region. In summary, the aging effect of journals facilitates the Bradford curve to more easily adopt an S-shape.

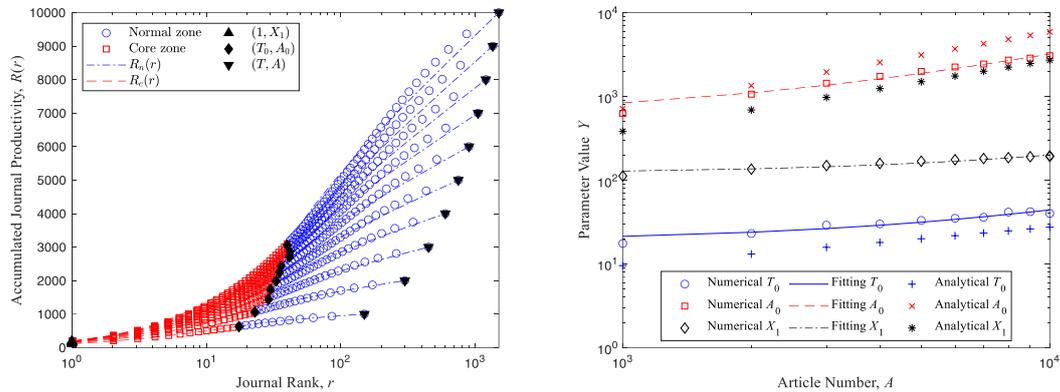

figure 6 the dynamics of the Bradford curve and the variation of key parameters when $\alpha = 0.15$ and $\gamma = 0.95$: (a) the dynamics of the Bradford curves; (b) the variations of key parameters.

### 3.3 Varying Entry and Aging Rates

Figure 7 shows the impact of varying entry and aging rates on Bradford curves. In real-world scenarios, both the entry rate and the aging rate often changes steadily. For example, the entry rate might decrease linearly from 0.2 to 0.1, while the aging rate increases linearly from 0.95 to 1.0, and the simulation results are shown in Figure 7. Comparing Figures 5, 6, and 7 reveals that the effects of decreasing entry rate and increasing aging rate are similar to those observed with constant entry and aging rate (Figure 6). However, in Figure 7(b), both the article number $A_0$ and journal number $T_0$ of are even lower compared to Figure 6(b), indicating that the decreasing entry rate further exacerbates their decrease. Consequently, the normal region of Bradford curve becomes less concave downward, and its starting point on the y-axis is notably lower. Importantly, all three key factors continue to show linear relationships with the article number, suggesting that Equation (18) remains useful for predicting them.

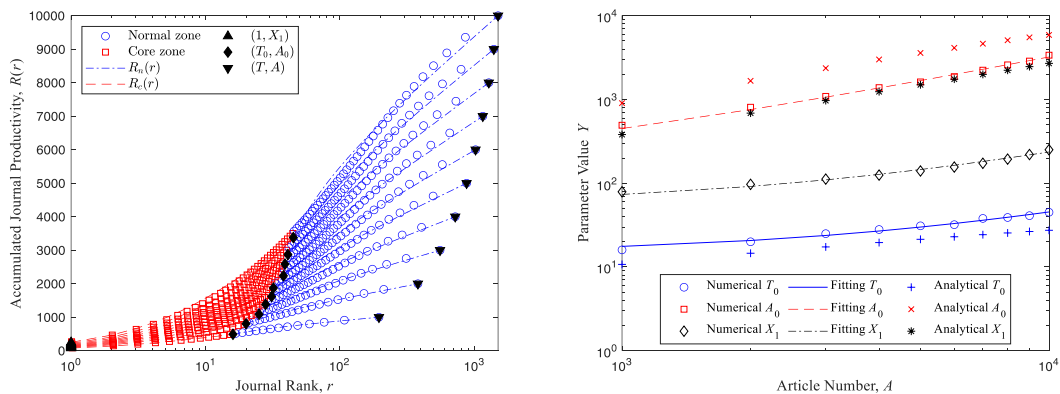

Figure 7 the dynamics of the Bradford curve and the variation of key parameters when $\alpha$ decreases linearly from 0.2 to 0.1 and $\gamma$ increases linearly from 0.95 to 1.0. (a) the dynamics of the Bradford curves; (b) the variations of key parameters.

### 4 Empirical Study

## 4.1 Dataset of Croatian Chemistry Research

Oluić-Vuković used the research output in chemistry by authors from Croatia to prepare full bibliographic references for a ten-year period (Oluić-Vuković, 1992). This dataset includes only articles published in journals, comprising 2,543 papers across 416 journals over a decade. The productivity of the top few (fewer than 10) most prolific journals was taken directly from Figure 1 in (Oluić-Vuković, 1992), while the productivity of other journals was taken from Tables 4 and 6 in (Oluić-Vuković, 1998). Data from the figures were adjusted to match the total number of journals and articles provided in Table 2 of (Oluić-Vuković, 1992).

To predict the dynamics of Bradford's curve, the first step is to predict the variation of the total article number $A(t)$ over time $t$. Logistic regression analysis (Verhulst, 1838) was applied to the empirical data to predict the total article number $A(t)$ at any given time, as shown in Figure 8(a). Next, the total journal number $T$ and the entry rate of new journals $\alpha$ were estimated by plotting the total journal number $T$ against the total article number $A$ and applying a linear fit of $T = A\alpha$, as shown in Figure 8(b). Once the point $(T, A)$ is determined for any time, linear regression of Equation (18) is used to determine the three key parameters $T_0$, $A_0$ and $X_1$ on a log-log axis, with the fitting results shown in Figure 9(a). This allows the key points $(T_0, A_0)$ and $(1, X_1)$ to be determined at any time. Finally, based on these three key points, Equations (8) and (13) can be used to determine the Bradford curve for the core region and the normal region, respectively, with results shown as red dashed lines and blue dotted lines in Figure 9(b).

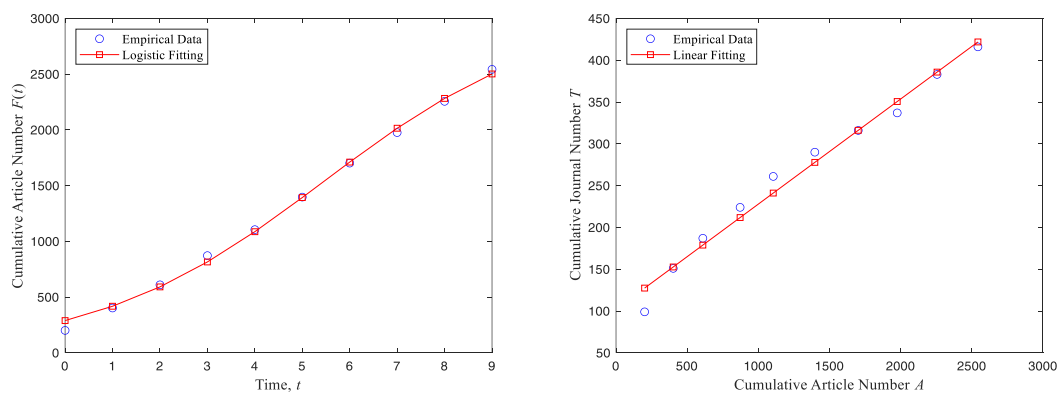

Figure 8 The process of determining the point $(T, A)$ for any given time. (a) the total article number $A(t)$ as a function of time $t$; (b) the total journal number $T$ as a function of the article number $A$.

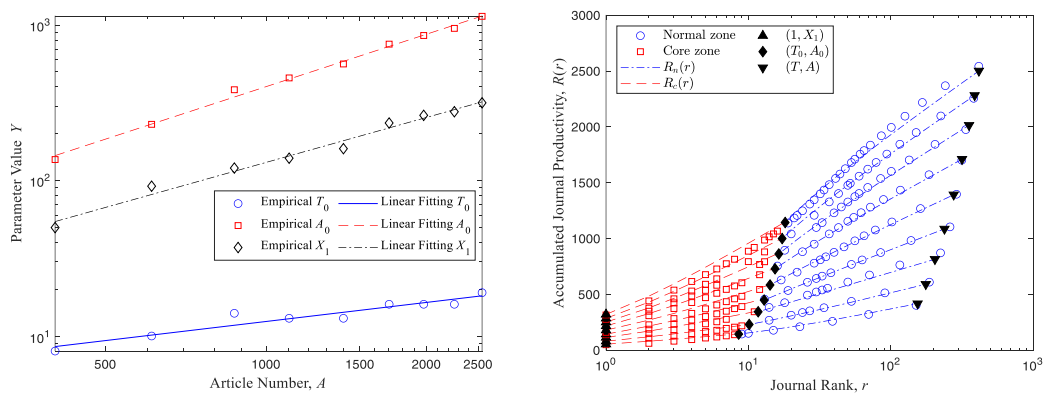

Figure 9 The procedures for predicting the evolution of Bradford curve. (a) the variations of key parameters $T_0$, $A_0$ and $X_1$ with the article number $A$; (b) the dynamics of the Bradford curves.

The Bradford curves shown in Figure 9(b) align closely with the empirical data, demonstrating a strong match between the predicted and observed outcomes. Notably, the Bradford curve transitions gradually from a J-shape to an S-shape, a transformation that is accurately captured by the analytical predictions.

**4.2 Dataset of Solar Power Research**

The bibliographies on solar power research for the years 1971, 1974, 1977, 1980, 1983, and 1986, encompassing papers published in journals from the Engineering Index, were compiled by Garg et al. (Garg et al., 1993). The data for this analysis were directly extracted from Tables 1-7 of the referenced study.

Similar to the Croatian Chemistry Dataset, predicting the dynamics of Bradford's curve involves the following four steps:

1. Predicting Article Number $A$: Apply logistic regression to the empirical data of cumulative article numbers $A(t)$ over time $t$ (from Table 7) to predict article numbers for the desired intervals, as shown in Figure 10(a).

2. Predicting Journal Number $T$: Apply the quadratic fitting of Equation (20) to the journal and article pairs to estimate the journal number $T$ and entry rate of new journals $\alpha$ given article number $A$, as illustrated in Figure 10(b).

3. Predicting key parameters $T_0$, $A_0$ and $X_1$: Use the linear fitting of Equation (18) on empirical data of $T_0$, $A_0$ and $X_1$ in the log-log axis to obtain their predictions at any article number $A$, as depicted in Figure 11(a).

4. Drawing Bradford Curve: Apply Equations (8) and (13) to draw the Bradford's curve for the core region and the normal region, respectively, as shown in Figure 11(b).

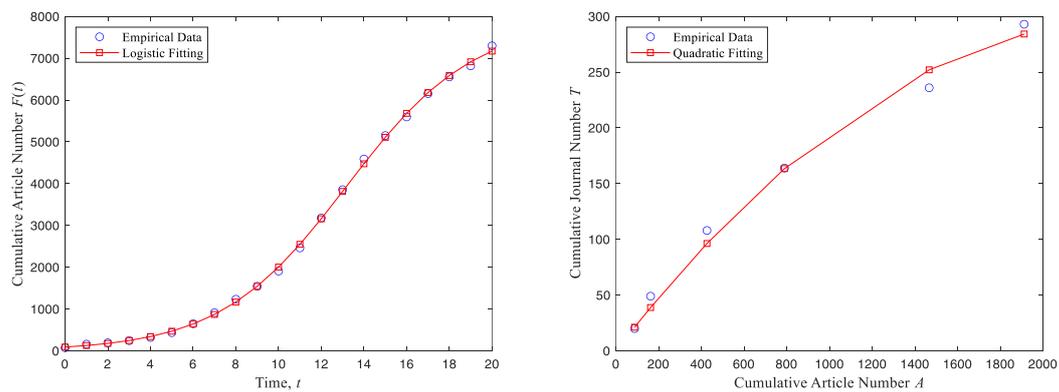

Figure 10 the process of determining the point $(T, A)$ for any given time. (a) the total article number $A(t)$ as a function of time $t$; (b) the total journal number $T$ as a function of the article number $A$.

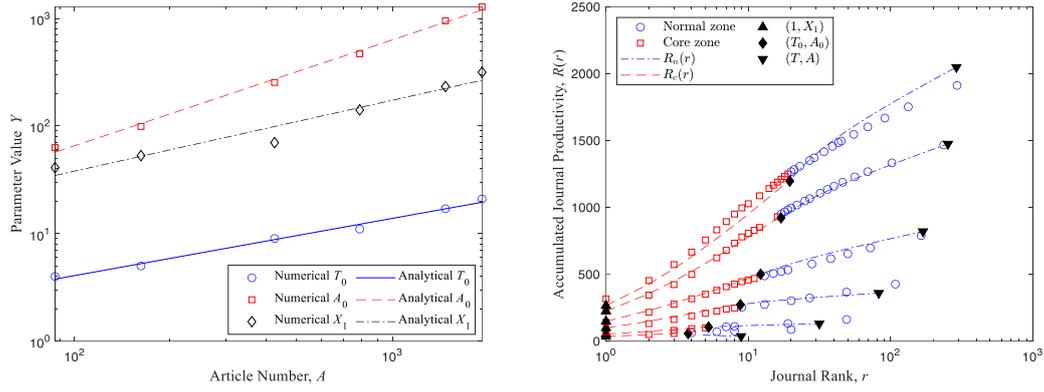

Figure 11 the procedures for predicting the evolution of Bradford curve. (a) the variations of key parameters $T_0$, $A_0$ and $X_1$ with the article number $A$; (b) the dynamics of the Bradford curves.

Figures 9(b) and 11(b) demonstrate that while the proposed method can predict the general trend of Bradford's curves, there are inherent errors. These errors arise because Bradford's law inherently contains uncertainties. Numerical studies reveal that the article numbers for each journal rank have a large standard deviation, making it practically impossible to predict the precise shape of Bradford's curve. Additionally, the various fitting procedures introduce errors into the process. Therefore, this method can only predict the general trend of Bradford's dynamics, but cannot accurately predict the article number for each journal at any given time.

Another issue with this method is that the first derivatives of the core region (Equation (14)) and the normal region (Equation (16)) differ at the $(T_0, A_0)$ point, resulting in a non-smooth analytical curve at the intersection point. In contrast, the numerical simulation results are smooth throughout. This problem could be addressed by proposing more complex formulas for the normal region, but this would complicate the overall method. Given the difficulty in accurately predicting Bradford's curve dynamics, this aspect is not explored further in this paper.

## 5 Conclusion

This paper examines how integer constraints on the number of journals $T$ and articles $A$ affect the shape of Bradford's curve, dividing it into two distinct zones: the core zone and the normal zone, based on the significance of these integer effects. Using the Simon-Yule model, we derive analytical results for key parameters and distributions under a constant entry rate. Theoretical formulas for each zone are developed, and the reasons behind the various shapes of Bradford's curves are analyzed. Monte Carlo simulations are employed to study the impact of decreasing entry rates of new journals and aging rates of journals on the shape of Bradford's curve and key parameters. Finally, we validate our proposed method using empirical data from Croatian Chemistry and Solar Power research datasets. The main conclusions are:

1. Bradford's curve should be divided into two separate zones based on the significance of integer constraints on journal and article numbers. Different formulas for each zone should be derived separately.

2. Bradford's curve can exhibit four different shapes, determined by the second derivatives of the core and normal zones.

3. The largest productivity $X_1$, the number of journals $T_0$, and the number of articles $A_0$ are key parameters influencing the shapes of Bradford's curves. Decreasing entry rates and aging rates of journals affect these parameters.

4. The proposed four-step method can predict general trends in Bradford's curves despite some errors.

The conclusions of this paper can guide academic libraries in procuring and utilizing scientific literature effectively.

## 6. Statements and Declarations:

**Funding and Conflicts of interests:** The research leading to these results received funding from the Library Society Guangdong under Grant Agreement No. GDTK23004. This article is one of the achievements of the 2023 Key Research Project of Guangdong Provincial Library titled "Joint Analysis and Data Governance of Papers and Patents in the Context of Smart Libraries" (Project Number: GDTK23004).

## 7. Author contributions:

All authors contributed to the study conception and design. Material preparation, data collection and analysis were performed by Haobai Xue and Xian Liu. The first draft of the manuscript was written by Haobai Xue and all authors commented on previous versions of the manuscript. All authors read and approved the final manuscript.